# Toward Ethical Spatial Analysis: Addressing Endogenous Bias Through Visual Analytics

Chuan Chen[a], Peng Luo[b]*, Bo Zhao[c], Yu Feng[a] and Liqiu Meng[a]

[a]*Chair of Cartography and Visual Analytics, Technical University of Munich, Munich, Germany;*

[b]*Senseable City Lab, Massachusetts Institute of Technology, Cambridge, USA;*

[c]*Department of Geography, University of Washington, Seattle, USA*

Spatial analysis can generate both exogenous and endogenous biases, which will lead to ethics issues. Exogenous biases arise from external factors or environments and are unrelated to internal operating mechanisms, while endogenous biases stem from internal processes or technologies. Although much attention has been given to exogenous biases, endogenous biases in spatial analysis have been largely overlooked, and a comprehensive methodology for addressing them is yet to be developed. To tackle this challenge, we propose that visual analytics can play a key role in understanding geographic data and improving the interpretation of analytical results. In this study, we conducted a preliminary investigation using various visualization techniques to explore endogenous biases. Our findings demonstrate the potentials of visual analytics to uncover hidden biases and identify associated issues. Additionally, we synthesized these visualization strategies into a framework that approximates a method for detecting endogenous biases. Through this work, we advocate for the integration of visualization at three critical stages of spatial analysis in order to minimize errors, address ethical concerns, and reduce misinterpretations associated with endogenous biases.

# 1. Introduction

Spatial analysis has become deeply integrated into human society, playing a pivotal role in driving innovation and enhancing efficiency across multiple critical sectors. Discussions about ethical considerations are constantly and existing research has already identified ethical issues in various domains, particularly those induced by bias [1][2]. This highlights the critical importance of addressing bias. Transferring the definition of bias to the geographical information science (GIS) context, geo-bias refers to systematic distortion or imbalance in geospatial data, modeling, or interpretation caused by geographic factors. However, systematic research on uncovering geo-bias remains scarce. One major reason is the absence of a well-defined methodology for systematically categorizing and analyzing geographic biases.

Bias can be broadly categorized into two primary types: exogenous bias, which emerge from potential misuse or unintended applications, and endogenous bias, which are inherent to the technology's design and underlying mechanisms. In the context of GIS research, exogenous bias results from misuse or external influences, such as misuse of GIS application. Endogenous bias, on the other hand, results from the inherent limitations of spatial data, spatial algorithms, or spatial workflow design, such as incorrect projections or oversimplified spatial models. Despite extensive research on the bias associated with the misuse of GIS, a significant gap exists in examining the endogenous bias embedded within the technology itself. Most current discussions are outcome-oriented, focusing predominantly on the consequences of misuse rather than addressing the underlying mechanisms that could render GIS applications ethically problematic. Consequently, the motivation for addressing endogenous biases has emerged.

Visualization has long been regarded as a crucial method for addressing bias [3][4].

However, there is a lack of specific studies demonstrating how visualization can be utilized to address endogenous biases in spatial analysis. In our study, we leverage the advantages of visual analytics, then employ it as an addressing tool to conduct endogenous bias addressing in spatial analysis. By deconstructing spatial analysis, the main components are extracted and organized into a coherent framework. For each components, the potential forms of endogenous bias and the corresponding visual analysis scrutiny methods have been identified and proposed, which ensures the practicality of the review process.

**2. Related work**

Geo-bias has been sparingly used in GIS studies, and even within the limited studies, interpretations and understandings vary. Some researchers employ it to investigate statistical paradoxes [5], while studies in spatial statistics define misinterpretation as a form of bias [6]. Even though there has already been extensive discussion on bias in the field of AI ethics with a coherent and comprehensive perspective [7][8], the GIS field urgently requires a well-defined framework and classification system to help address this issue comprehensively. After categorizing bias into endogenous and exogenous types, exogenous bias has received most attention. Several studies have investigated the misuse of geographic data for unauthorized tracking and surveillance, as well as the potential for unequal distribution of social resources due to biased or inequitable use of data [9,10,11,12]. Besides the data misuse, another typical ethical issue arising from exogenous bias or the misuse of GIS applications is privacy. One research emphasizes the potential risks to privacy posed by geospatial artificial intelligence technologies [13]. To address privacy more comprehensively, a concept termed "collective privacy," distinct from individual privacy, has been introduced, highlighting privacy risks at the group or community level [14]. Additionally, issues related to the reuse of volunteered geographic data raise further privacy conflicts. Specific data formats have been proposed as a solution to safeguard privacy in these contexts [15]. Collectively, these investigations have not only highlighted critical ethical risks especially including bias but have also elevated standards and expectations for the responsible use of GIS technology, introducing the need for robust frameworks to address these risks in spatial analysis.

While biases from GIS misuse are well-studied, endogenous biases within the technology and their ethical implications remain largely overlooked. Although issues of bias and fairness have been widely explored from a statistical perspective [5,13], these analyses often overlook the foundational sources of bias intrinsic to GIS technology. This gap shows the necessity of identifying factors that could cause GIS application results ethically questionable, even under conditions where external use is conducted in a fully correct and ethical manner. In this context, endogenous bias denotes ethical challenges that arise directly from the technology's design and application, independent of user intentions or actions. As a result, there is a growing imperative to investigate how endogenous bias might compromise the ethical integrity of GIS technology, leading to a re-evaluation of spatial analysis practices. In other words, the motivation for addressing endogenous bias in spatial analysis has emerged.

However, endogenous biases are often difficult to uncover. This is because a lack of understanding of the data and the model itself makes it challenging to detect issues in the results. In the GIS field, this can lead to erroneous spatial decisions that go unnoticed by decision-makers, ultimately resulting in unfair treatment of individuals in their daily lives. In recent years, among the research for addressing bias, some studies have proposed applying visualization techniques to address the issue[3][4], providing the potential of visual analytics in addressing the endogenous bias in GIS [16]. In the field of data management, visual analytics offers advantages in its auditing and scrutiny functions

[17][18]. In the field of geography, visual analytics has been proven effective in uncovering potential data-generating processes and identifying relationships among variables [19]. Therefore, we think it can also play a significant role in addressing endogenous ethical issues in GIS which mainly arises from the improper understanding of spatial data and the inherent data-generating processes. However, to the best of our knowledge, there is no existing study exploring the potential of visual analytics in detecting the endogenous ethical bias in GIS.

## 3. Endogenous bias in spatial analysis

Spatial analysis refers to the process of modeling geographic data and interpreting the results to draw conclusions, which normally include three primary components: data, modeling, and interpretation [20]. Each of these components possesses distinct characteristics that shape the spatial analysis process, while simultaneously interacting with one another to influence the overall outcomes of the analysis.

First, data serves as the foundation of spatial analysis, recognized early on for its unique characteristics, particularly spatial autocorrelation and spatial heterogeneity [20][21]. These distinct properties contribute to the critical role of data in spatial analysis, influencing both the methods used and the insights generated. The significance of data has been widely acknowledged, with its impact on the future development of spatial analysis frequently emphasized [22][23].

Second, modeling forms the another core component of spatial analysis. Initially, the primary function of spatial modeling was to simulate geographic phenomena, thereby providing insights into complex spatial interactions [24]. Over time, spatial models have come to be regarded as essential tools within spatial analysis, facilitating the transformation of raw spatial data into meaningful interpretations that support decision-making and understanding [25][26].

Finally, model interpretation is closely linked to the results of spatial analysis. Its role was one of the earliest aspects of spatial analysis to receive scholarly attention, particularly within exploratory spatial data analysis [27][28]. As spatial analysis techniques have become increasingly applied across diverse fields, interpretation has gained importance in elucidating spatial patterns relevant to various disciplines, such as, disease mapping, ecology, archaeology, and other fields of social and environmental sciences [29,30,31,32].

To address endogenous bias systematically, we selected representative spatial analysis workflows originating from these three core elements in spatial analysis: data, modeling, and interpretation as illustrated in Figure 1.

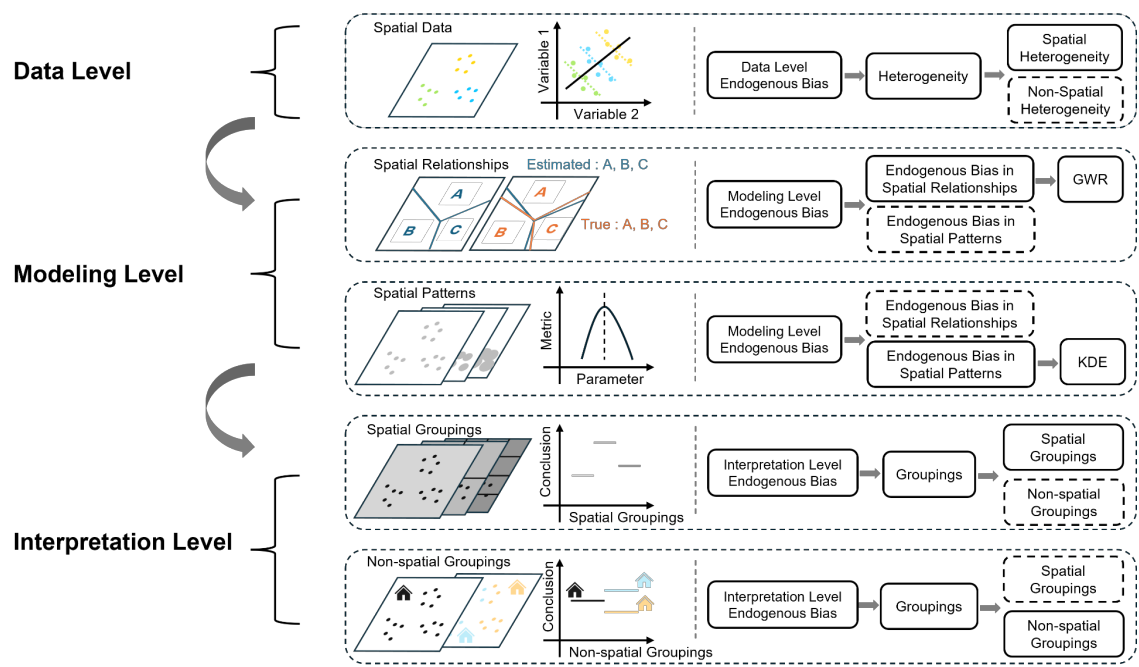

Figure 1: Representative spatial analysis workflows selected to examine endogenous biases

At the data level, spatial data exhibits many unique characteristics that are primarily related to widely discussed spatial effects, heterogeneity and dependence [20][21]. Spatial dependence, to some extent, can assist in modeling, as seen in techniques like kriging interpolation [33]. In the absence of sufficient data or explanatory variables, dependence can provide valuable information. In contrast, heterogeneity poses major challenges to spatial analysis.

In statistic, such heterogeneity can significantly impact the accuracy and interpretability of analytical results [34]. For example, a notable consequence of data heterogeneity is Simpson's Paradox, a well-documented phenomenon in statistics [35]. Simpson's Paradox occurs when trends observed within grouped subsets of data are inconsistent with trends in the overall dataset, potentially leading to wrong conclusions, as illustrated in Figure 2. It represents this paradox by showing multiple clusters of data points, each with a distinct color representing a different subgroup. Within each subgroup, the trend, represented by a dashed line, shows a negative relationship between variables $u$ and $v$. However, when the data for all subgroups is aggregated, the overall trend, represented by the red dashed line, reverses, indicating a positive relationship between $u$ and $v$. This figure illustrates how ignoring heterogeneity across data dimensions can lead to conflicting interpretations, emphasizing the importance of addressing heterogeneity to ensure accurate spatial analysis.

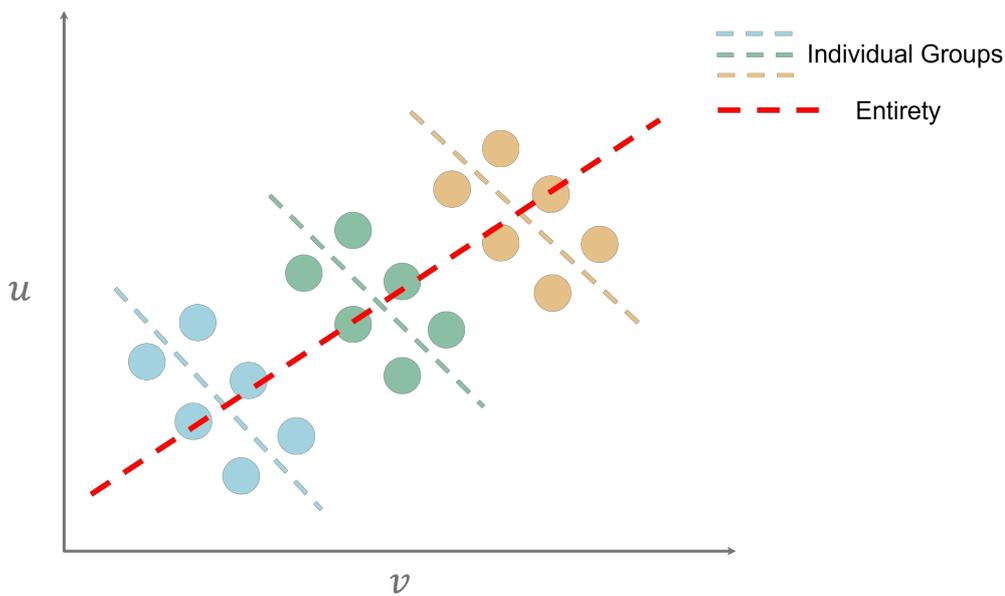

Figure 2: Schematic of Simpson's Paradox

At the modeling level, an inadequate understanding of the generative processes of geographic data can lead to incorrect or inappropriate modeling approaches. Taking two important spatial analysis tasks, the analysis of spatial relationships and the extraction of spatial patterns, as examples, we demonstrate how modeling-level biases arise. First, when identifying spatial correlations, most models have a strong assumption of linearity such as geographically weighted regression (GWR) [36][37]. This becomes an important factor contributing to endogenous bias as shown in Figure 3. Extraction errors in regions with nonlinear spatial relationships can impact policy-making by misrepresenting spatial boundaries and resource distributions. As illustrated in the figure, Policy 1 and Policy 2 are initially represented as separate areas with clear-cut boundaries. However, due to extraction errors of spatial relationships, an overlapping region (represented by the hatched area) emerges, creating ambiguity in policy jurisdiction and implementation. This overlap can result in policy conflicts, where resources and responsibilities are either duplicated or overlooked, leading to inefficiencies and misaligned outcomes.

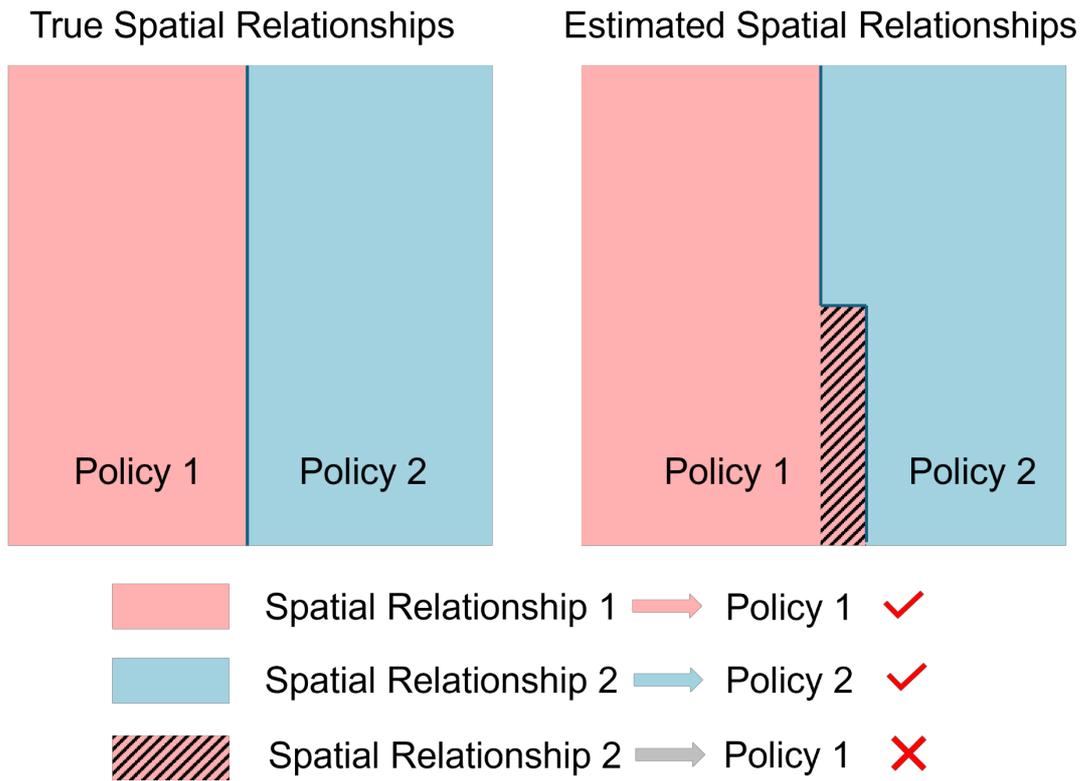

Figure 3: Schematic of endogenous bias in spatial relationships

In the context of spatial pattern extraction, parameter selection plays a crucial role in determining the accuracy and reliability of extracted patterns. Unreasonable or inadequately filtered parameter choices can introduce various forms of endogenous bias, leading to uncertainty. For instance, in the commonly applied Kernel Density Estimation (KDE) method, parameters such as bandwidth and sample size are critical to achieving accurate results [38][39]. Inappropriate choices of these parameters can drastically alter the outcome, potentially leading to significant errors or even conclusions that are entirely contrary to reality.

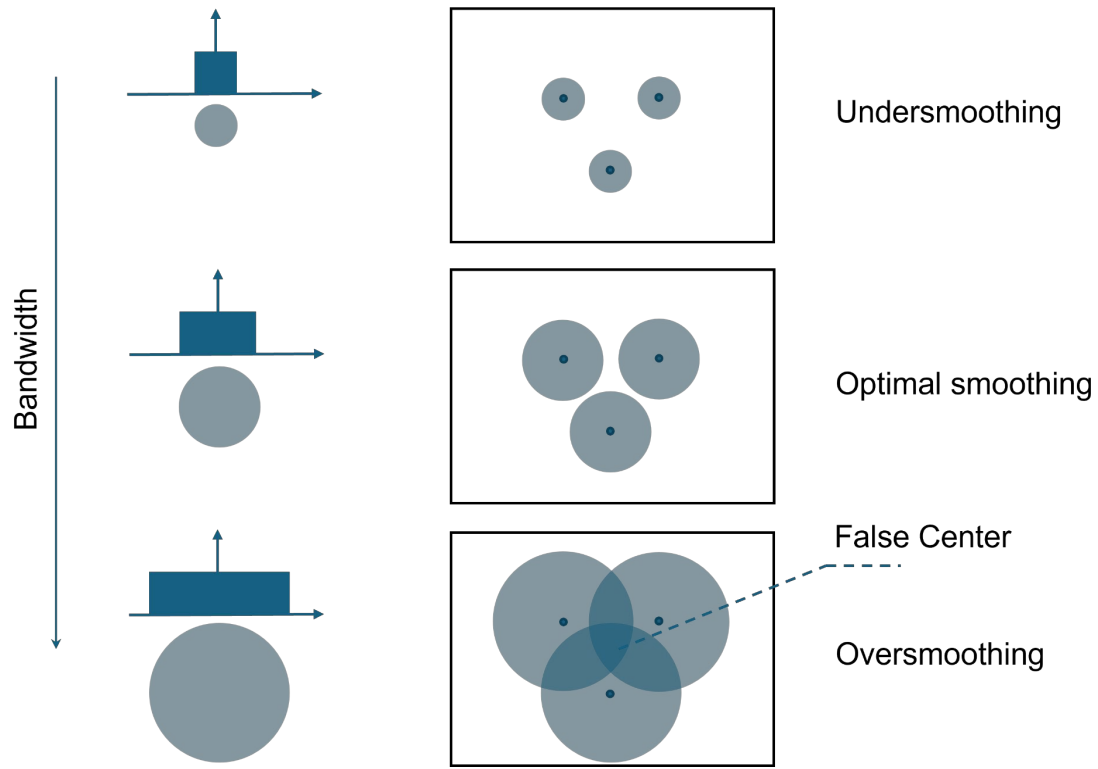

Figure 4: Schematic of endogenous bias in spatial patterns

As shown in Figure 4, different bandwidth parameters yield different results. When the bandwidth is too large, a "false center" phenomenon may occur, where the focus of spatial pattern appears in an originally low-density data area.

At the interpretation level, it is necessary to aggregate and model the data based on the initially selected analytical units or groups, and then interpret the modeling results. Consequently, the choice of different units or groups may influence the conclusions of the analysis. Grouping is therefore a direct factor affecting the final results and may lead to endogenous bias.

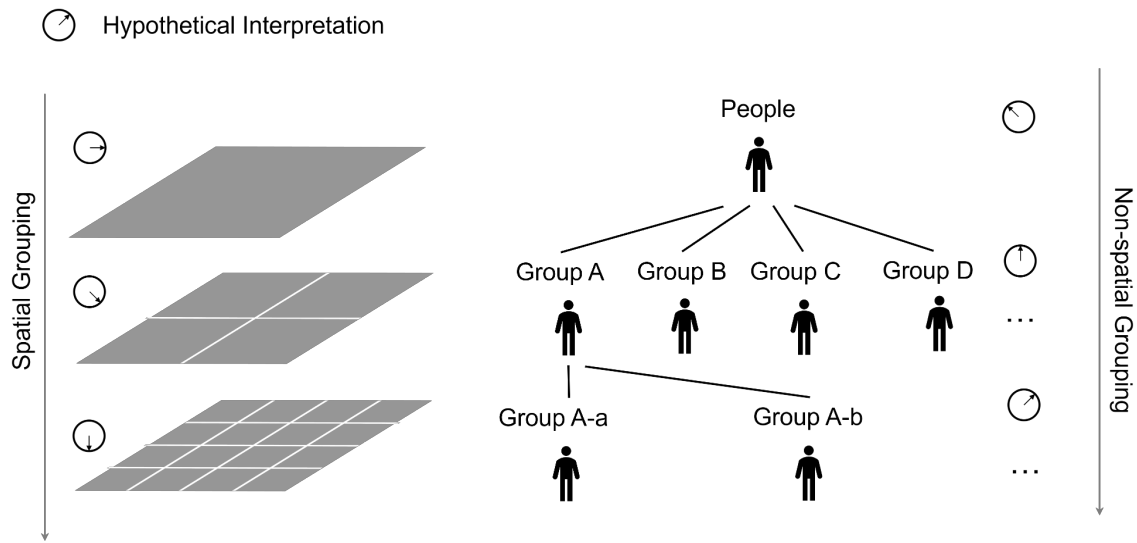

Figure 5: Schematic of spatial and non-spatial grouping

The concept of grouping here includes not only spatial grouping but also non-spatial grouping of the data as shown in Figure 5. The Modifiable Areal Unit Problem (MAUP) is a widely discussed issue that illustrates how spatial grouping can affect interpretation outcomes [6]. Similarly, the grouping issue also exists in non-spatial data. This research will illustrate a case of endogenous bias that arises from the division of demographic data into racial groups.

**4. Visual analytics as a tool to address endogenous bias**

Visualization is an effective tool for detecting and examining bias [40][41]. It allows researchers to intuitively identify patterns, inconsistencies, and biases within complex datasets. However, existing research lacks the systematic development of visualization strategies to demonstrate the process of addressing bias. In this study, simulated and real-world data will be used to test and describe how visual analytics can address endogenous bias in spatial analysis. We will demonstrate it at three levels: data, modeling, and interpretation as shown in Figure 6.

To effectively address the endogenous bias of data heterogeneity in spatial analysis at the data level, we propose the use of dimensionality reduction, an effective method for detecting and handling heterogeneous data [42], combined with visualization to detect heterogeneity. At the modelling level, we propose spatial continuity tests of model parameters to scrutinize this form of endogenous bias. We then propose dynamic parameter visualization for model evaluation to reveal the endogenous bias in spatial patterns. Finally, at the interpretation level, we propose multi-grouping visualization to reveal endogenous bias caused by spatial grouping or non-spatial grouping.

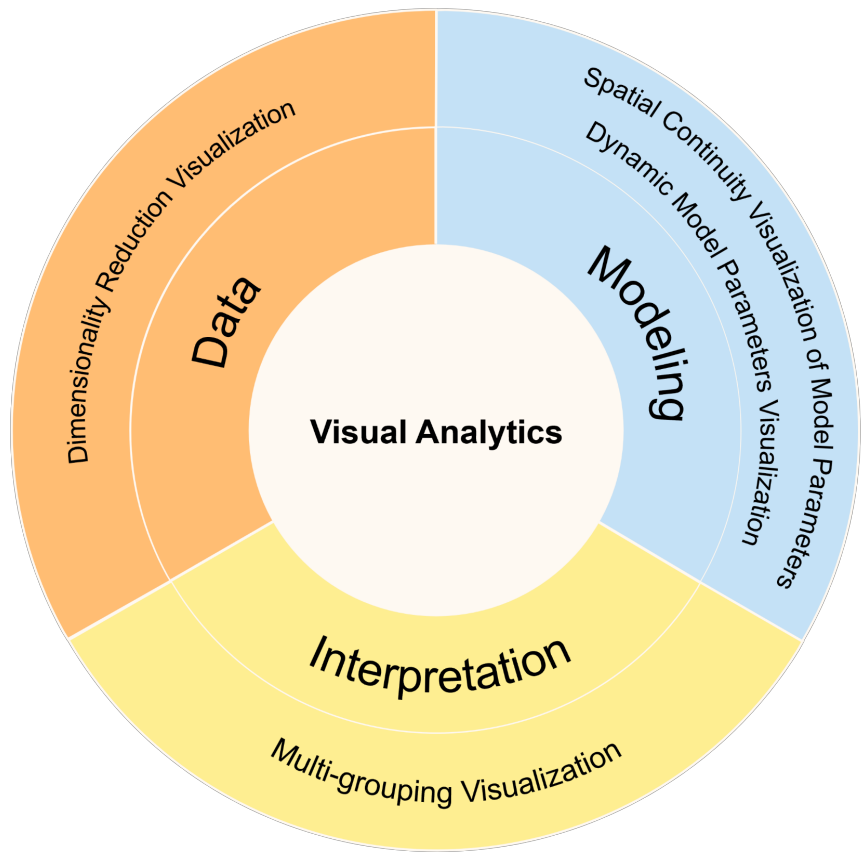

Figure 6: A three-tiered strategy for visual analytics that reveals endogenous bias

## 4.1 Data-level endogenous bias

### 4.1.1 Endogenous bias caused by heterogeneity

To examine the bias induced by spatial heterogeneity at the data level, we designed an experiment that includes three heterogeneous regions (A, B, and C), within which we simulated the occurrence of Simpson's Paradox in spatial analysis, as illustrated in Figure 7.

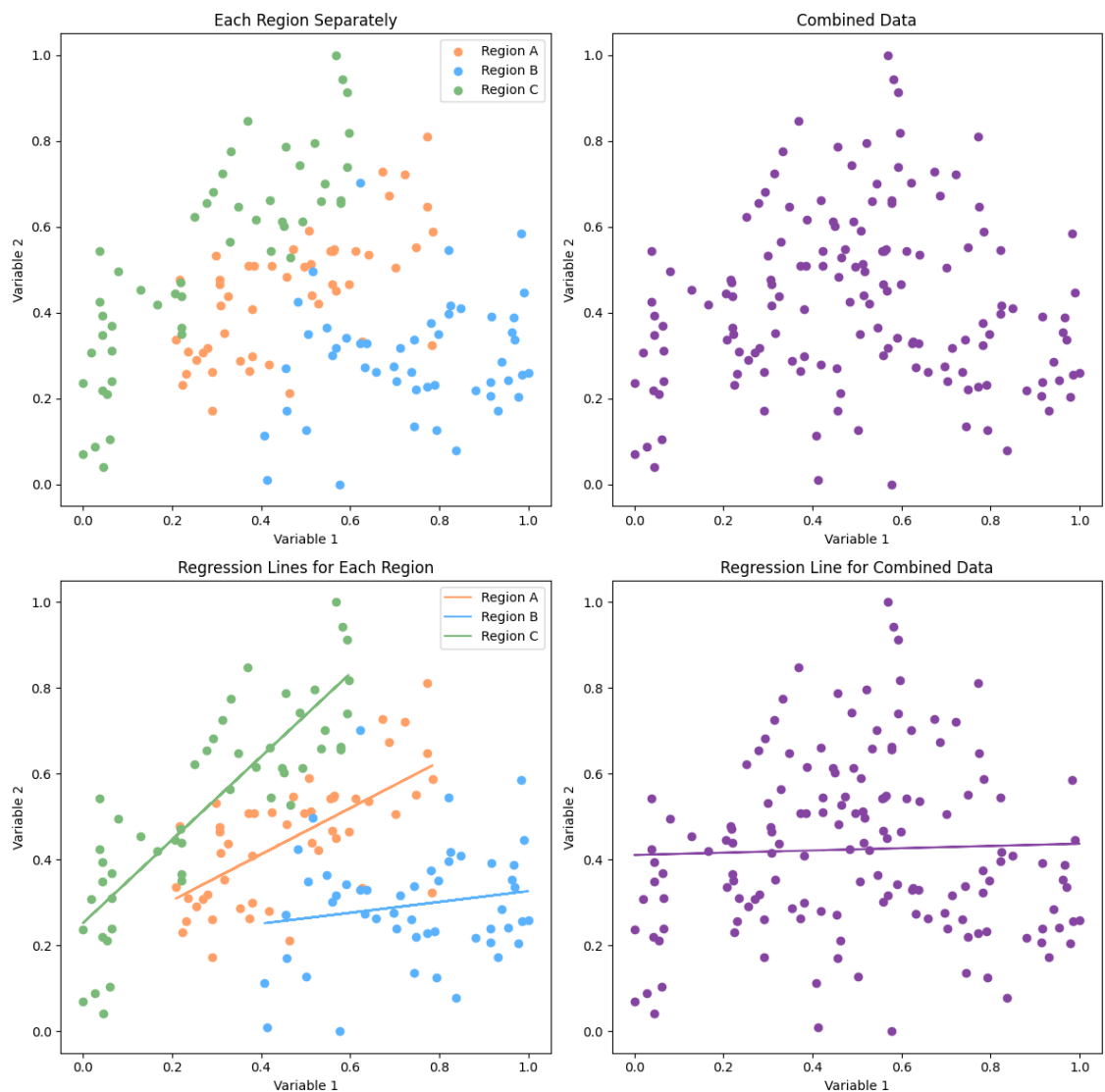

Figure 7: The Simpson's Paradox due to spatial heterogeneity

In this experiment, Variable 1 and Variable 2 are set to exhibit a positive correlation within each of the three regions. Both grouped regression and overall regression analyses are performed on the data from these regions to evaluate the consistency of correlations. In the grouped regression analysis, as expected, Variables 1 and 2 display a positive correlation in each region, consistent with the predefined conditions. However, in the overall regression analysis that considers data from all regions, these two variables do not demonstrate a significant correlation. This discrepancy indicates an instance of Simpson's Paradox, which can be regarded as an example of endogenous bias.

By leveraging visual analytics, dimensionality reduction techniques were applied to further examine the heterogeneity in the data. Parallel coordinate plots were used as an effective means of visual dimensionality reduction [43]. Figure 8 presents a visualization of the simulated data with two variables and two spatial coordinates as four parallel coordinates, where a clear grouping effect is visible, highlighting the heterogeneity among regions. Treating regions A, B, and C as one single entity in the analysis of the correlation between Variable 1 and Variable 2 obviously constitutes endogenous bias, as the spatial heterogeneity is ignored.

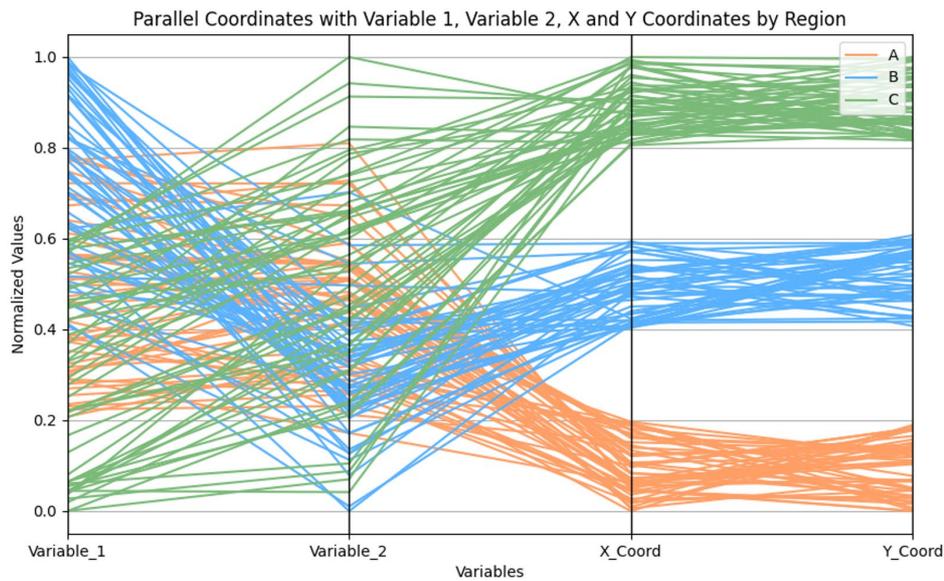

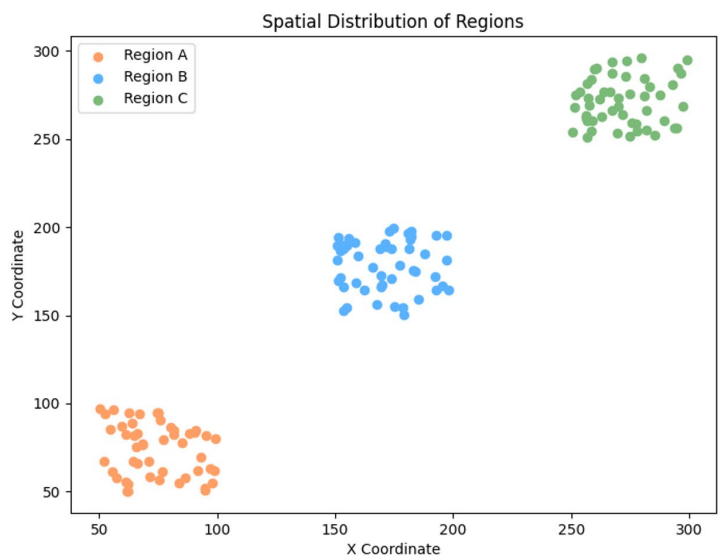

Figure 8: Dimensionality reduction and mapping of simulated data

If this heterogeneity is not addressed at the data level, the resulting conclusions may be fundamentally misleading. What makes this issue particularly worrying is that neglecting heterogeneity does not necessarily make the entire analysis process appear flawed. The data may still appear to be correctly analyzed, but it can potentially lead to erroneous conclusions within what appears to be an accurate spatial analysis process. In other words, such errors are subtle and often difficult to detect. To this end, the visual analytics with the combined approach of data dimensionality reduction and visualization helps identify the hidden errors or instances of endogenous bias.

*4.2 Modeling-level endogenous bias*

In modeling level, endogenous bias may mainly arise from two primary sources. The first source is the neglect of examining the spatial continuity of spatial parameters during the extraction of spatial relationships. The second is the lack of dynamic parameter visualization when identifying spatial patterns. We designed experiments to simulate how the visual analytics can effectively reveal instances of endogenous bias during the modeling process.

*4.2.1 Endogenous bias in spatial relationship*

GWR is one of the most common methods for extracting spatial relationships. The core assumption of GWR is strong linearity, meaning that specific relationships or parameters are expected to be linear within the spatial regression model. However, in many cases, the relationship between independent and dependent variables may deviate from linearity, even on a local scale. In regions where spatial parameters exhibit discontinuity, GWR results can contain significant errors, leading to endogenous bias.

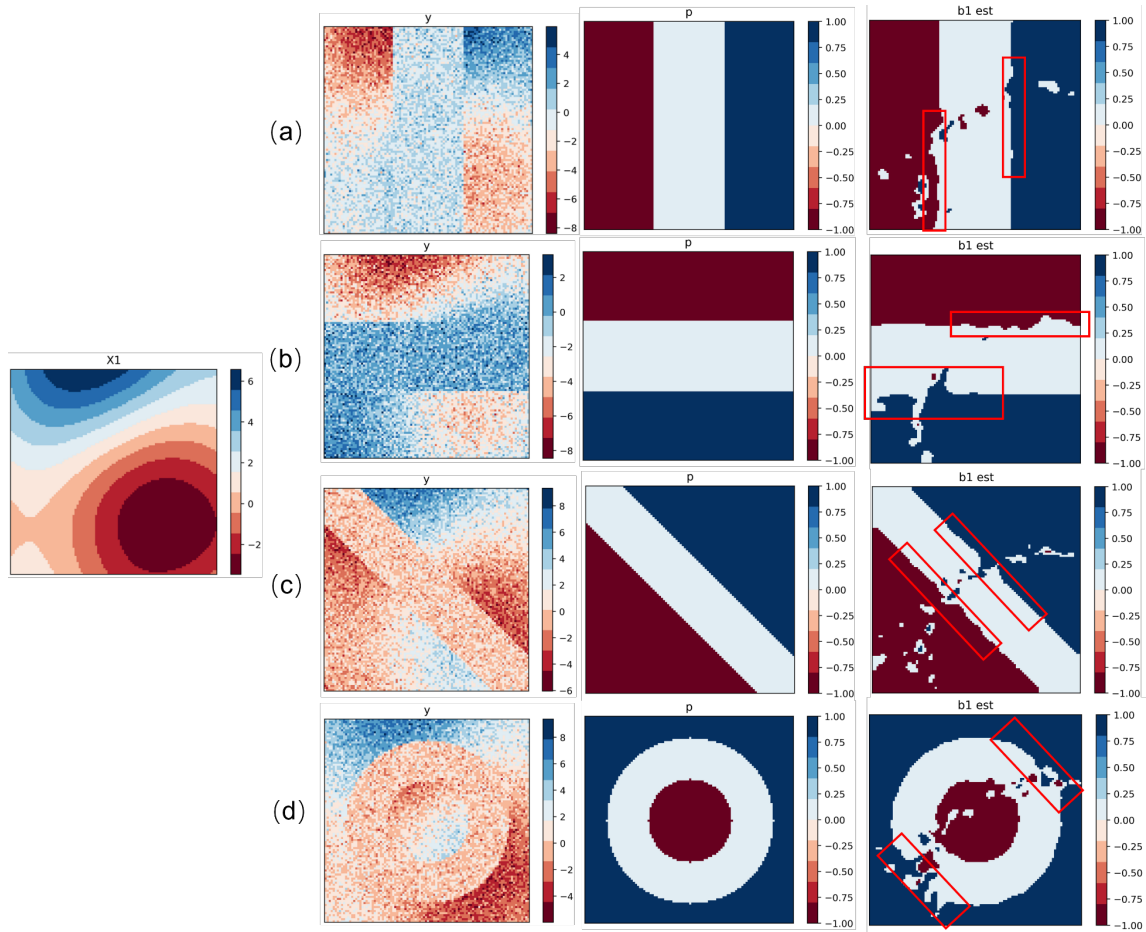

Figure 9: Four visualization results show the limitations of GWR in handling nonlinear spatial relationships.

To investigate this issue, we designed 4 experiments to demonstrate how the visual analytics helps identify and reveal this form of endogenous bias, as illustrated in Figure 9. In these simulations, a set of predefined spatial relationships is expressed by the following formula:

$$y = X1 \times p + error$$

where $y$ represents the dependent variable and $X1$ denotes the independent variable. $p$ represents the parameter to be fitted. The variable $b1\ est$ is the fitted result of GWR and post-optimization processing. Spatial continuity testing is conducted on $b1\ est$ as shown in Figure 9. The spatial distribution of the parameter $b1\ est$ demonstrates regions where discontinuities occur. The red boxes highlight the errors caused by the discontinuities in the spatial parameters. The results indicate that areas with spatial discontinuities in *b1 est* estimates tend to exhibit a high concentration of errors in the GWR model. In other words, errors in the GWR fitting process tend to cluster in regions with spatial parameter discontinuity, a manifestation of endogenous bias. Using visual analytics, continuity testing of spatial parameters, as applied to *b1 est*, enables the detection of such errors. This approach completes the scrutiny of endogenous bias by pinpointing areas where the GWR model assumptions are violated, thereby enhancing the reliability of spatial analysis.

### *4.2.2 Endogenous bias in spatial patterns*

When extracting spatial patterns, inappropriate parameter settings can lead to endogenous bias. Dynamic visual analysis of parameters is therefore useful. Taking the bandwidth of KDE as an example of parameter setting, we designed two experiments to study the impact of bandwidth on the results of KDE, thus the accuracy of spatial pattern extraction [44][45].

In the first experiment, we set up a dataset showing a local low-density spatial pattern along with a dataset showing a global spatial pattern which contains the local dataset as shown in Figure 10. KDEs are performed on the two datasets at different scales to extract the spatial patterns. Subsequently, a local window is selected, and the gradient directions of the KDE results from both instances within this window are displayed. The bandwidths for the two experiments are set to 0.6070 and 0.3526, respectively, to achieve optimal spatial pattern recognition, based on Silverman's rule of thumb [46]. The results reveal significant deviations in local gradient directions due to the different bandwidth choices, regarded as the endogenous bias.

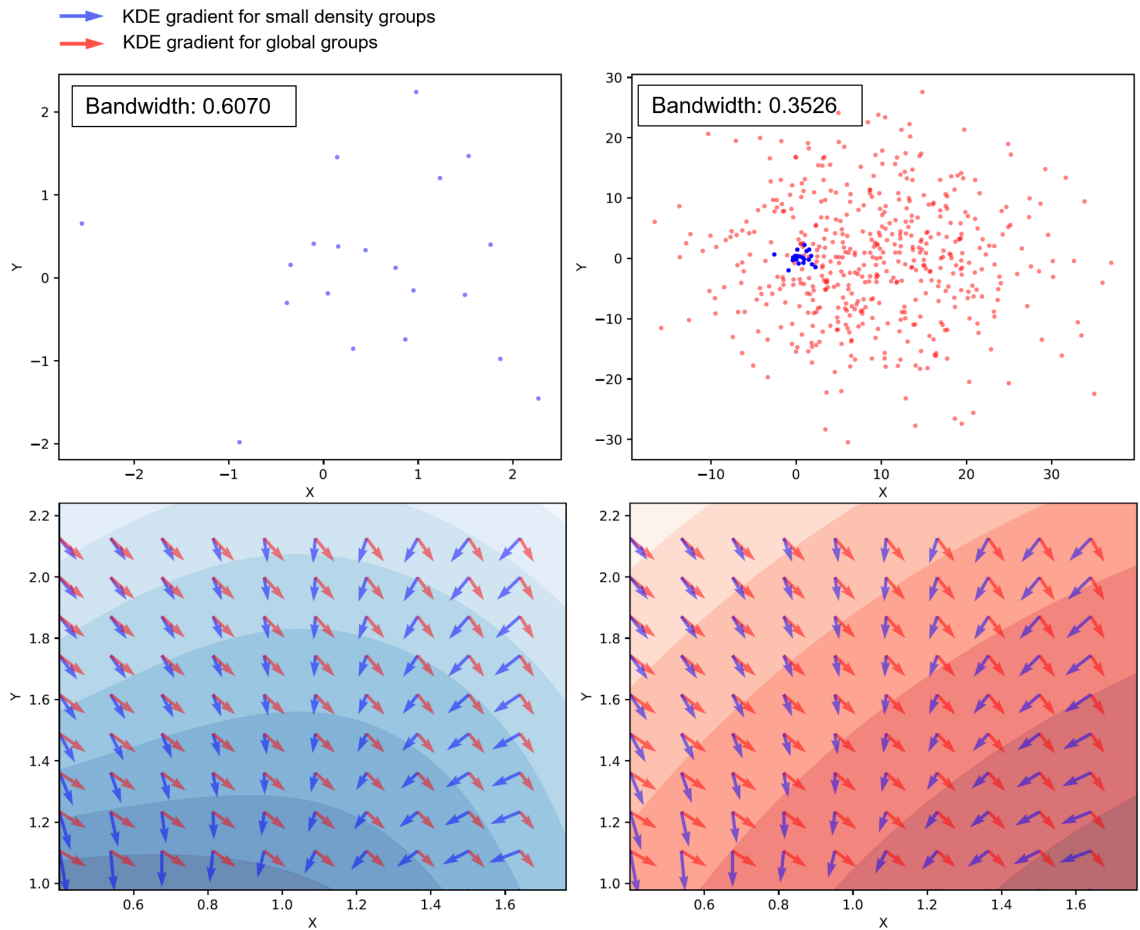

Figure 10: Impacts of bandwidth selection on model outcomes

Furthermore, the second simulation experiment is conducted, and the KDE results of the spatial pattern distribution of three clustered groups with continuously varying bandwidth are displayed, as shown in the Figure 11.

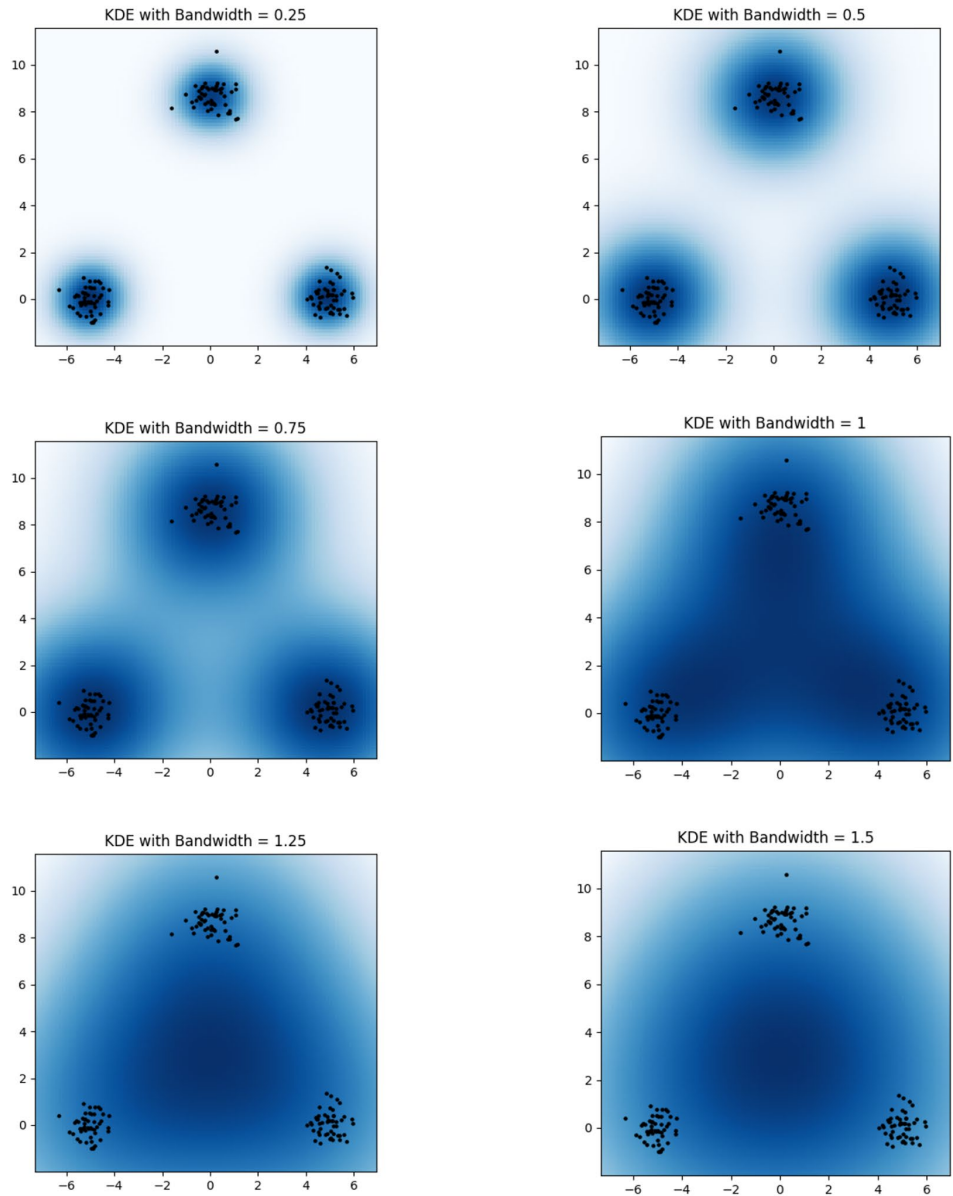

Figure 11: The visualization of dynamic bandwidths

From this experiment, it can be observed that spatial patterns can only be accurately extracted when the bandwidth is within an appropriate range. In cases where the bandwidth is excessively large, even the center of the distribution may be incorrectly represented, appearing in locations that originally have very low density.

In both of the above experiments, it is shown that inappropriate spatial parameters can lead to endogenous bias. Through the dynamic parameter visualization, the impact of bandwidth on the results is intuitively demonstrated, which can help improve the awareness of endogenous bias.

*4.3 Interpretation-level endogenous bias*

At the interpretation level, the grouping of georeferenced statistical data can significantly influence interpretation outcomes, potentially leading to endogenous bias. One widely discussed issue related to spatial grouping is the Modifiable Areal Unit Problem (MAUP), which shows how different spatial groupings can yield varying results. Moreover, interpretation grouping covers both spatial and non-spatial groupings and both can affect analysis outcomes. This section is dedicated to a simulation experiment and an experiment using a real-world dataset with the aim to demonstrate how multi-grouping visualization can help identify and mitigate the MAUP problem in interpretation.

## 4.3.1 Endogenous bias caused by spatial grouping

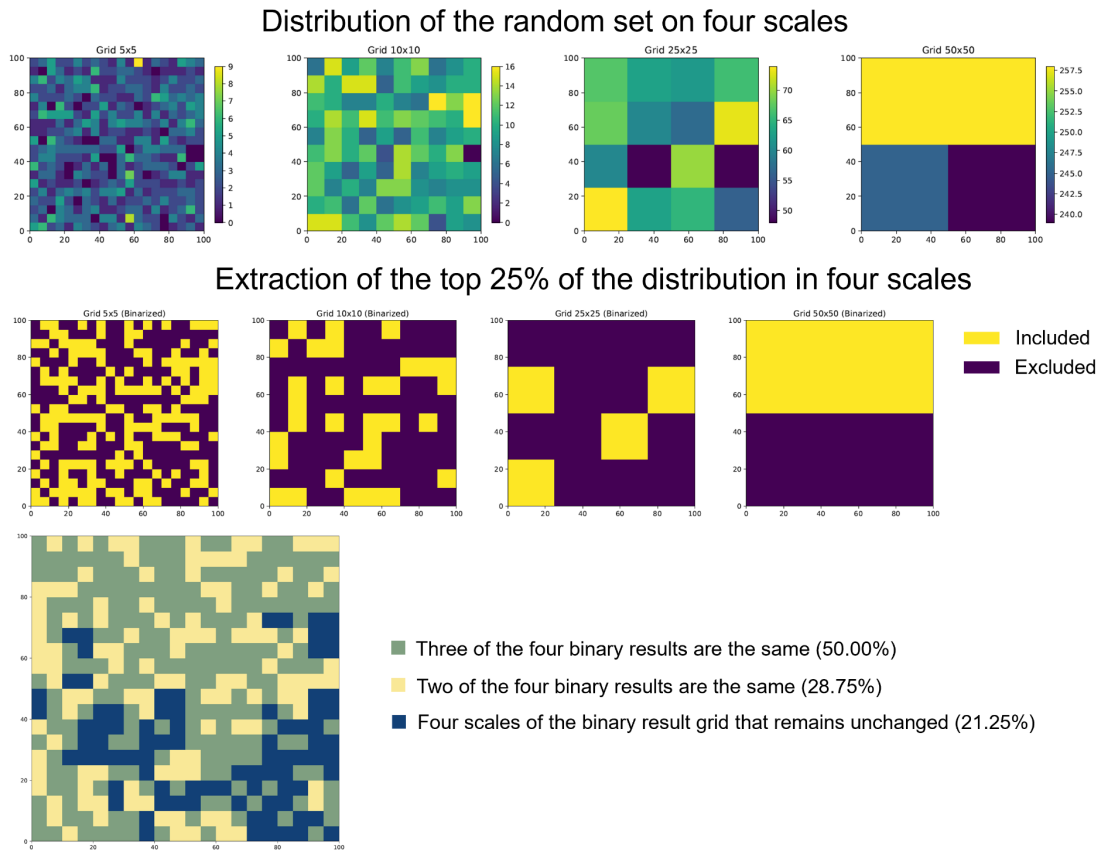

Figure 12: Interpretation influenced by MAUP

A simulation experiment is conducted to investigate the endogenous bias caused by varying spatial groupings, as shown in Figure 12. After randomly constructing a spatial distribution with a side length of 100, an analysis task is set up to extract the regions where the attribute values are within the top 25%. The results of the analysis task are visualized at four different spatial groupings. There are obviously significant discrepancies in the unified spatial analysis interpretation across groupings. Furthermore, we conduct a statistical analysis of the consistency of the interpretation results using a grid with a side length of 10.

In Figure 12, the results reveal that only 21.25% of the grids maintain consistent binarization classification across all four spatial groupings. In contrast, 28.75% of the grids display consistency for only two groupings, while the remaining 50.00% of the grids exhibit varying classifications across one or more of the four groupings. These findings reveal the substantial impact that changes in grouping can have on analytical conclusions, potentially leading to endogenous bias. The visual analytics of different groupings can facilitate a detailed examination of how grouping discrepancies influence spatial interpretations and contribute to endogenous bias.

*4.3.2 Endogenous bias caused by non-spatial grouping*

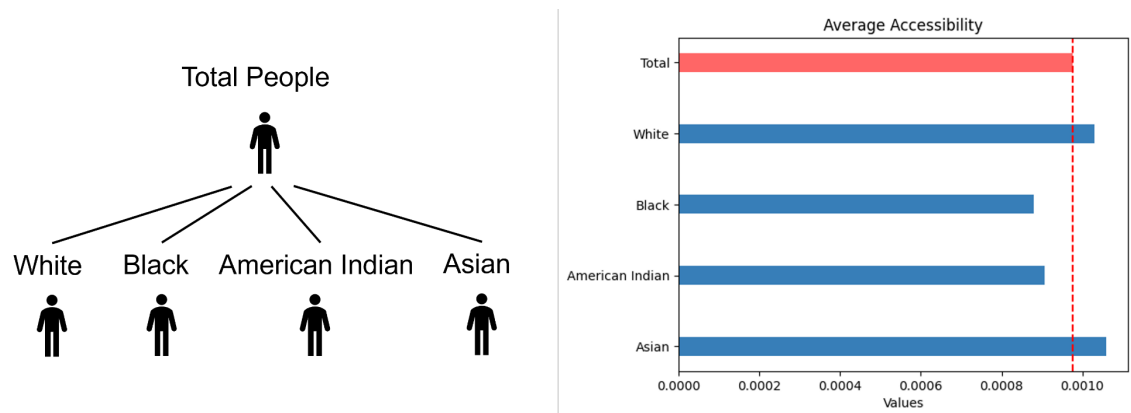

Figure 13: Average hospital accessibility in Cook County according to ethnic groups

Similar to Simpson's paradox caused by spatial heterogeneity of geographic variables, conducting different grouping methods of geographic analysis results and drawing conclusions based on these groupings also implies endogenous bias. For instance, as th analysis of population data becomes more refined, it becomes possible to distinguish between different ethnic groups or to further subdivide the population into more detailed demographic categories. This shift in grouping methods can result in varying interpretations and potentially lead to wrong conclusions, underscoring the need to account for both spatial and non-spatial groupings to ensure responsible and accurate data interpretation.

We conducted an accessibility analysis using hospital and ethnic data from Cook County, Chicago [47][48]. The hospital accessibility for the total population is calculated by a method of three-step floating catchment area access (3SFCA) as shown in Figure 13 [49]. Combined with the ethnic data, the average hospital accessibility for different ethnic groups is presented. Based on different groupings of population data, the results of the accessibility index are also visualized.

When considering the entire population as an entity, the overall accessibility score is found to be 0.000977. When the grouping is conducted, the data reveal that White individuals have an accessibility score of 0.00103, indicating slightly better access compared to the overall average. In contrast, Black individuals have a lower score of 0.000879, suggesting they face more barriers in accessing hospital services. Similarly, the American Indian population has a score of 0.000907, also below the overall average. Asian individuals, however, have the highest accessibility score at 0.00106, reflecting the best access among the racial groups analyzed. The significant differences in healthcare accessibility among different ethnic groups highlight the inconsistency between optimization objectives at large-grouping and small-grouping population data levels. This endogenous bias arises when the overall optimization goals for a region do not align with the localized optimization needs of specific subregions or communities. In this case, while county-wide planning and resource allocation aim to maximize overall accessibility, they inadvertently neglect the unique requirements and challenges faced by individual racial groups. As a result, certain populations, such as Black and American Indian communities, experience lower accessibility scores compared to others.

This misalignment between non-spatial, global strategies and local realities shows the need for more nuanced, targeted approaches in healthcare planning to ensure equitable access for all racial groups within the county.

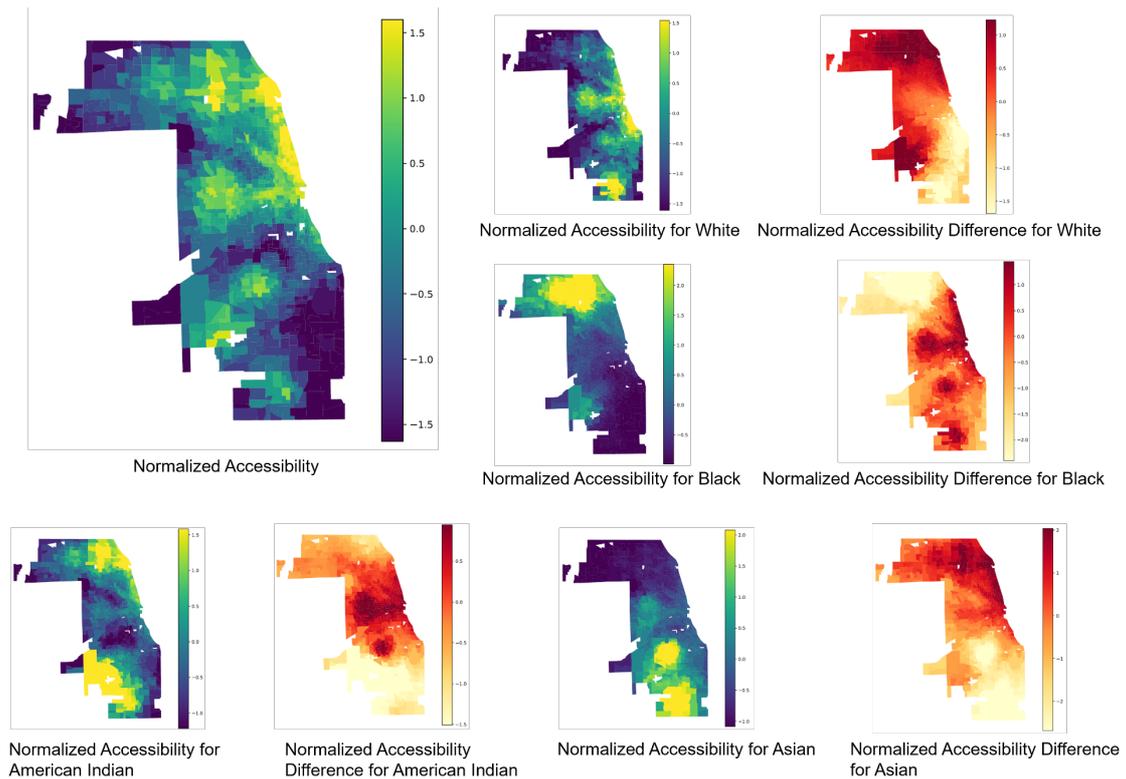

Figure 14: Normalized distribution of overall and different ethnic group accessibility and their difference display

In addition to calculating hospital accessibility for different ethnic groups, the normalized distribution of hospital accessibility for each group is also visualized as shown in figure 14. The hospital accessibility distribution presents a markedly different picture when considering only specific ethnic groups. For Whites, hospital accessibility is higher in the central and southern areas when only Whites are considered. However, the difference results indicate that the overall optimization strategy decreases hospital accessibility in the southern part of Cook County for Whites, as evidenced by the low difference values. When focusing on Blacks, their hospital accessibility is higher in the northern area. Yet, the difference results reveal that the overall optimization strategy significantly reduces hospital accessibility in the north for Blacks. Similar patterns are observed for the American Indian and Asian groups. The results of this experiment demonstrate that spatial interpretation outcomes can vary significantly under different non-spatial data groupings. By applying visual analytics, we may effectively highlight this form of endogenous bias. The framework allows for a deeper understanding of how overall optimization strategies can overlook the unique needs of specific communities, addressing disparities that may otherwise remain hidden in aggregated data. This insight emphasizes the importance of incorporating multi-grouping analysis to address the distinct accessibility challenges faced by different racial groups.

## 5. Concluding remarks

This study categorizes ethics issues in GIS into two types: exogenous and endogenous. The focus is on the often-overlooked endogenous bias, which are analyzed through the lens of the spatial analysis process. The sources of these biases are identified in such process levels: data, modeling, and interpretation. Since these sources are deeply embedded throughout the spatial analysis, they are frequently go unnoticed. However, they significantly arise from a lack of understanding of the unique characteristics of spatial data, which can result in flawed modeling and interpretation.

In this work, we explore the potential of visual analytics as a tool to identify endogenous bias in spatial analysis. While visualization is widely recognized for its role in detecting irresponsibility in data science, its application to addressing endogenous bias in spatial analysis remains underutilized. To bridge this gap, we propose leveraging visual analytics to systematically uncover endogenous bias in spatial analysis. By transforming analytical results into intuitive graphical formats such as maps and charts, visual analytics facilitates the detection of patterns, trends, and anomalies, making it an indispensable approach for mitigating endogenous bias in spatial analysis.

We selected case studies addressing the three primary sources of endogenous bias in spatial analysis and examined them to explore the role of visual analytics. Our analysis highlights that dimensionality reduction visualization is particularly effective in uncovering endogenous bias stemming from data heterogeneity. Additionally, visualizing both the spatial continuity and dynamic changes of model parameters helps to expose endogenous bias introduced during modeling. For biases emerging during interpretation, multi-grouping visualization proves to be a powerful tool for identifying and addressing such issues.

In summary, we propose the toward ethical spatial analysis approach through visual analytics, leveraging visualization techniques to uncover endogenous biases in spatial analysis. This visual analytics framework enables the effective detection of such biases across most spatial analysis tasks. In other words, our study provides a preliminary approach to identifying endogenous biases and their subsequent ethical risks in spatial analysis tasks.

**Data and codes availability statement**

The data and codes that support the findings of the present study are available on Figshare at https://figshare.com/s/9b896174e9dee14ce0d4.

**Disclosure Statement**

No conflict of interest exists in this manuscript, and the manuscript was approved by all authors for publication.